\documentclass{cmmp}
\usepackage{graphics}
\let\realverbatim=\verbatim
\let\realendverbatim=\endverbatim
\renewcommand\verbatim{\par\addvspace{6pt plus 2pt minus 1pt}\realverbatim}
\renewcommand\endverbatim{\realendverbatim\addvspace{6pt plus 2pt minus 1pt}}

\renewcommand\thesection{\arabic{section}}
\newcommand{\roughly}[1]{\raise.3ex\hbox{$#1$\kern-.75em
\lower1ex\hbox{$\sim$}}}

\begin{document}
\setcounter{chapter}{1}

\hskip 4 in\vbox{\baselineskip12pt\hbox{hep-th/0205027}\hbox{SU-ITP-02/17}}

\bigskip
\bigskip
\bigskip
\centerline{\Large\bf Black holes at accelerators}
\bigskip
\bigskip
\centerline{{\bf Steven B. Giddings}\footnote{email address:  
giddings@physics.ucsb.edu.  }}
\bigskip
\centerline{
Department of Physics, University of
California, Santa Barbara, CA\ 93106}
\bigskip
\centerline{and}
\bigskip
\smallskip
\centerline{Department of Physics and SLAC,
Stanford University, Stanford,
CA\ 94305/94309}
\bigskip
\bigskip
\bigskip
\bigskip
\bigskip
\centerline{\bf Abstract}
\bigskip
\noindent
This is my contribution to the proceedings of Stephen Hawking's 60th birthday celebration.  If the ideas of TeV scale gravity are correct, then black holes should be produced at accelerators that probe the TeV scale, and their decays should evidence one of Stephen's greatest discoveries, the phenomenon of black hole radiance.

\chapter*{Black holes at accelerators}

\section{Introduction}

I'd like to begin with a birthday wish for Stephen's birthday $\ldots$ in 2008.
It comes in the form of an email message from the future Director of CERN,
dated January 8, 2008, which reads:
\begin{verbatim}
Dear Professor Hawking,

We wish to alert you to an announcement that will be made at 
a press conference tomorrow.  Since the recent start up of 
LHC, both ATLAS and CMS have seen numerous events with
large numbers of jets and hard leptons, large transverse 
momentum, and high sphericity.  These are consistent with 
TeV-scale black hole production, and in particular with 
extrapolations of your predictions for black hole radiance 
to higher dimensions.  The press conference is timed to 
coincide with publication of the results in Phys. Rev. Lett.  


With best regards,

Director General, CERN

\end{verbatim}
It is supplemented with a present for the rest of us:
\begin{verbatim}
PS
You may also be interested to know that there appear to be 
anomalous correlations and other very interesting structure 
hinting at resonances at the high-energy end of the 
spectrum of decay energies.  Further details will appear in 
the forthcoming paper.
\end{verbatim}
The point of my talk is to explain that, with an optimistic view of
TeV-scale gravity scenarios, such an amazing development could become a
reality.  (In more pessimistic scenarios, we have to build a higher energy
machine!)

What might we hope to learn from this turn of events?  First, we'd be able
to experimentally confront head on a profound problem that Stephen lead us
to with his discovery of black hole radiance\cite{Hawking:sw}:  the black hole information
paradox.  At the same time, we might also learn a lot of other things about
the quantum mechanics of gravity, and perhaps experimentally confirm the
ideas of string theory.  This would be more than we've ever dared to hope
for.

In outline, I'll start by describing some of the basic ideas of TeV-scale
gravity, which make this remarkable scenario feasible.  I'll then turn to a
description of black holes on brane worlds and their production in high-energy 
collisions.  Next is a discussion of black hole decay, where
Stephen's calculations come to the fore.  I'll close by describing some of
the other consequences of this scenario, including what appears to be the
end of investigation of short-distance physics, but may be the beginning of
the exploration of the extra dimensions of space.  For a more in-depth
treatment of the subject of black hole production in TeV-scale gravity (and
more complete references), the
reader should consult the original references:  
\cite{Giddings:2001bu} for the overall story (see also
\cite{Dimopoulos:2001hw}); the more recent paper \cite{Eardley:2002re},
which treats the classical problem of black-hole formation in high-energy
collisions, and \cite{Giddings:2001ih}, which serves as another review,
with further discussion of black hole creation in cosmic ray collisions
with the upper atmosphere\cite{Feng:2001ib,DGRT,Anchordoqui:2001ei,Emparan:2001kf,Ringwald:2002vk}

\section{TeV-scale gravity}

The idea that the fundamental Planck scale could be as low as the TeV scale
is the essential new idea that inevitably leads to black hole production at
energies above this threshhold.  TeV-scale gravity is a novel approach to the
long-standing {\it hierarchy problem}:  we observe what appear to be two
centrally important scales in physics, the weak scale $M_W\sim 1 TeV$,
and the four-dimensional Planck scale, $M_4\sim \frac{1}{\sqrt G}\sim 10^{19}
GeV$, where $G$ is Newton's gravitational constant; 
what explains
the huge ratio between them?  Traditional views invoke supersymmetry and
its breaking, but the new idea is that the {\it fundamental} scale in
physics is the TeV scale, and that the observed weakness of gravity,
corresponding to the high value of $M_4$, results from dilution of
gravitational effects
in extra dimensions of space.

To explain further, suppose that there are $D$ total dimensions of
spacetime, with coordinates $x^\mu$, $\mu=0,1,2,3$ parametrizing the ones
we see, and $y^m$, $m=4,\ldots,D-4$ parametrizing the small 
compact ones we don't.  The most
general spacetime metric consistent with the very nearly Poincar\'e
invariant world we see is 
\begin{equation}
ds^2 = e^{2A(y)} \eta_{\mu\nu} dx^\mu dx^\nu + g_{mn}(y) dy^mdy^n\ ,\label{wmet}
\end{equation}
where $A$, conventionally called the ``warp factor,'' is an arbitrary
function of the unseen coordinates, $\eta_{\mu\nu}$ is the Minkowski
metric, and $g_{mn}$ is an arbitrary metric for the compact dimensions.
Gravitational dynamics is governed by the D-dimensional Einstein-Hilbert
action, and 
the action for four dimensional gravity is found by inserting
(\ref{wmet}), with a general four-dimensional metric $g_{\mu\nu}$, into this:
\begin{equation}
S_D= \frac{M_P^{D-2}}{4(2\pi)^{D-4}} \int d^D x \sqrt{-g} R \rightarrow \frac{M_4^2}{4} \int d^4 x
\sqrt{-g_4(x)} R_4 +\cdots\ .
\end{equation}
Here the extra terms on the right hand side are cancelled by whatever
matter lagrangian is necessary to make the metric (\ref{wmet}) a solution
to the D-dimensional Einstein equations.  
Define the 
`warped
volume'' of the extra dimensions,
\begin{equation}
V_W = \int d^{D-4} y \sqrt{g_{D-4}} e^{2A}\ .
\end{equation}
The critical equation is 
\begin{equation}
\frac{M_4^2}{M_P^2} = \frac{M_P^{D-4}}{(2\pi)^{D-4}} V_W\ :
\end{equation}
the ratio of the observed and fundamental Planck scales is given by the
warped volume in units of the fundamental Planck length.

We now have two options.  The first one is the conventional one: assume
that $M_P\sim M_4\sim 10^{19} GeV$, which means 
\begin{equation}
V_W\sim \frac{1}{M_P^{D-4}}\ .
\end{equation}
The new alternative arises if $M_P\sim 1 TeV \ll M_4$, and this can be
attained if the warped volume is for some reason very large:
\begin{equation}
V_W \gg \frac{1}{M_P^{D-4}} \sim \frac{1}{TeV^{D-4}}\ .
\end{equation}

There are two approaches to achieving this.  The first is the original idea of \cite{Arkani-Hamed:1998rs}:  simply take the volume to be large.
 A second approach is to take the warp factor to be large; a toy model of this type was introduced in \cite{Randall:1999ee}. An obvious objection then arises, which is particularly clear in the large volume scenario: the size of the extra dimensions ranges from around a millimeter for $D=6$ to  $10fm$ for $D=10$, and gauge interactions have already been tested far past this, to around $10^{-3}fm$, for example in the context of precise electroweak measurements.  Fortunately string theory has a made-to-order solution to this problem, which is the notion of a D-brane.  For example, suppose that there are six extra dimensions, but that there are some three-branes present within them; ordinary matter and gauge fields may be composed of open strings, whose ends are restricted to move in the three-dimensional space defined by the brane, whereas gravity, which is always transmitted by closed string exchange, will propagate in all of the dimensions.  String theory realizations of such ``brane-world" scenarios with large volume were described in \cite{Antoniadis:1998ig}, and string solutions with large warping were derived in \cite{Giddings:2001yu}.  We still lack completely realistic solutions with all the features to reproduce the known physics of the Standard Model at low energies, but ideas on this subject are still developing, or it may even be that such a scenario is realized outside of string theory.

\begin{figure}[t]
\scalebox{1.0}{\includegraphics{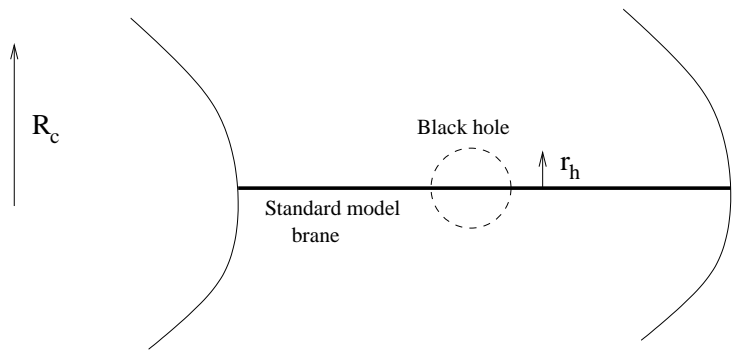}}
\caption{A schematic representation of a black hole on a brane world.  Gauge fields and matter are confined to the brane, but the black hole extends into all of the dimensions.  We consider the approximation where the black hole size is small as compared to characteristic geometric scales.}
\end{figure}

\section{Black holes on brane worlds}

Suppose, therefore, that we live on such a brane world.  As many
relativists have long expected, if we collide two particles, {\it e.g.} 
quarks, with
sufficiently high energy, they should form a large black hole; for a recent
discussion
of this see \cite{Banks:1999gd}.\footnote{For discussion of some properties
of black holes in higher-dimensional scenarios see 
\cite{Argyres:1998qn}; black hole decay was discussed in \cite{Emparan:2000rs}
and production 
of black holes in warped scenarios in \cite{Giddings:2000ay}.}
Being a
gravitational object, this black hole will extend off the brane, as
pictured in fig.~1.  Study of such black holes is greatly simplified
by using two approximations.  The first is to assume that the size of
the black hole, $r_h$, is much less than the size of the extra
dimensions or curvature scales of the extra dimensions, denoted $R_c$
in the figure.  This is typically true since the extra dimensions are
``large."  The second is the ``probe-brane" approximation: in general
the brane produces a gravitational field, but we neglect this field.
This is justified if the black hole is massive as compared to the
brane tension scale, which we expect to be approximately the Planck
mass.  So for large, but not too large, black holes, these two
approximations reduce the problem to that of describing solutions in
$D$ flat dimensions.

Black holes created in particle collisions will typically have some angular momentum; spinning black hole solutions in $D$ dimensions were first studied by Myers and Perry, in \cite{Myers:un}.  The horizon radius, Hawking temperature, and entropy of these black holes is given in terms of their mass $M$ and spin $J$; in the $J\rightarrow0$ limit, these take the form
\begin{equation}
\label{horrad}
r_h(M,J)\rightarrow \mathrm{constant} \cdot M^{\frac{1}{D-3}}\ ,
\end{equation}
\begin{equation}
T_H(M,J)\rightarrow \frac{\mathrm{constant}}{r_h}
\end{equation}
and
\begin{equation}
S_{BH}\rightarrow \mathrm{constant} \cdot M^{\frac{D-2}{D-3}}\ .
\end{equation}

We'd like to estimate the rate at which such black holes would be produced in high-energy accelerators.  The energy frontier is currently at proton machines.  At the fundamental level, proton collisions are collisions among their constituents, quarks and gluons, generically called partons.  In order to compute a rate, we need to know the density of partons of a given energy in the proton, or \textit{parton distribution function},
and the cross-section for a pair of partons to make a black hole.  Estimates of the parton distribution function are well known, but the black hole cross-section is not.  

The cross section can, however, be estimated by making a key observation: for large center-of-mass energies of the partons, the formation process should be essentially \textrm{classical}.  Indeed, consider a collision of two partons, each of energy $E/2$.  If they pass closely enough, we expect them to form a big black hole, with horizon radius determined by $E$.  For $E\gg M_P$, the curvature should be very weak at the black hole horizon, quantum gravity effects should be negligible, and a classical treatment should hold. 

An estimate of the cross section follows from Thorne's hoop conjecture\cite{Thorne:ji}, which suggests that if energy $E$  is concentrated in a region less than the corresponding Schwarzschild radius, $r_h(E)$, a black hole forms. If black holes form for impact parameters less than $r_h(E)$, this then indicates that that the cross section is given approximately by
\begin{equation}
\sigma\sim \pi r_h^2(E)\  .  \label{naicross}
\end{equation}

Indeed, recently Eardley and I have revisited the classical problem of black hole production in high energy collisions\cite{Eardley:2002re}, which was investigated some years ago for the case of zero impact parameter by Penrose\cite{Penrose} and D'Eath and Payne\cite{D'Eath:hb,D'Eath:hd,D'Eath:qu}.  In particular, Penrose found a closed-trapped surface in the geometry describing the head-on collision of two Aichelberg-Sexl solutions, and this implies that a black hole of mass at least $E/\sqrt2$ forms.  We extended this analysis to non-zero impact parameter.  In the case of four dimensions, we explicitly found that a trapped surface forms for impact parameter $b\roughly< 1.6 E$; this gives an estimated cross-section about $65\%$ the na\"{\i}ve value (\ref{naicross}).  We also find the area of the trapped surface, providing a lower bound on the mass of the resulting black hole.  While the $D>4$ problem has not been explicitly solved, we did reduce it to boundary-value problem for Poisson's equation, which corresponds to a small displacement soap bubble problem.  We expect this to yield similar results to those in $D=4$.  One can also explicitly see that the trapped surface forms before the collision -- and consists of two disks in the collision surface connected by a catenoid between them,\footnote{I thank L. Lindblom, M. Scheel, and  K. Thorne for conversations on this.  This construction has been subsequently rediscovered in \cite{Yoshino:2002br}.} as well as showing that the trapped surfaces can be deformed away from the curvature singularity at the center of the Aichelberg-Sexl solution\cite{Eardley:2002re}.  This builds a fairly convincing case for black hole formation, with a cross section not too far from the estimate (\ref{naicross}).  

Current experimental bounds\cite{Peskin:2000ti,Giudice:1998ck,Mirabelli:1998rt} on the Planck mass place it at $M_P\roughly>1.1-.8  TeV$ for $D=6-10$.  The LHC has a design center-of-mass energy of 14 TeV.  Suppose we make the optimistic assumption that indeed $M_P\sim 1$ TeV (and $D=10$).  One other important point is that one does not expect to create legitimate semiclassical black holes until one is a ways above the Planck scale, say at a minimum mass of $5-10 $ TeV.  The results are quite impressive\cite{Giddings:2001bu,Dimopoulos:2001hw}:  if the minimum mass for a black hole is 5 TeV, then LHC should produce black holes at the rate of about one \textit{per second}.  This would qualify LHC as a black hole factory, without bending standard nomenclature too far.  If the minimum mass is 10 TeV, then we'd still produce black holes at the rate of about three per day. 

Furthermore, from (\ref{naicross}) and (\ref{horrad}) we see that the cross-section for black hole production grows as
\begin{equation}
\label{ }
\sigma\sim E^{2/D-3}\ ,
\end{equation}
and so becomes even more dominant at higher energies.  

\section{Black hole decay and signatures}

The resulting black holes will decay in several phases.  When the horizon first forms, it will be very asymmetric and time dependent.  The first thing that should happen is that the black hole sheds its hair, in a phase we term \textit{balding}.  It does so by classically emitting gauge and gravitational radiation.
The amount of energy that can be emitted in this phase will be bounded given the bound on the minimum mass of the black hole resulting from the calculation of the area of the trapped surface.  This may be improved by perturbative methods, as was done by D'Eath and Payne in \cite{D'Eath:hb,D'Eath:hd,D'Eath:qu}.  A rough estimate is that somewhere between 15-40\% if the initial collision energy of the partons is shed in this phase.  The relevant decay time for this phase should be ${\cal O}(r_h)$.  We expect the black hole to rapidly lose any charge as well.
At the end of this phase we are left with a Kerr black hole with some mass and angular momentum.  

Next quantum emission becomes relevant, and Stephen's famous calculation
comes to the fore.  The black hole will Hawking radiate.  As shown by
Page\cite{Page:ki}, it first does so by preferentially radiating particles
in its equatorial plane, shedding its spin.  We call the corresponding
phase \textit{spindown}.  Page's four-dimensional calculations indicate
that about 25\% of the original mass of the black hole is lost during
spin-down; we might expect the higher-dimensional situation to be similar.
However, an important homework assignment is for someone to redo Page's
analysis of decay of a spinning black hole in the higher-dimensional
context, and fill in the details.\footnote{Since the conference, some
progress on this has been made in \cite{Kanti:2002nr}.}

At the end of spindown, a Schwarzschild black hole remains, and will
continue to evaporate through the \textit{Schwarzschild phase}.   This phase
ends when Stephen's caluculations fail, once the Schwarzschild radius
becomes comparable to the $D$-dimensional Planck length, and quantum
gravity effects become important.\footnote{In string theory there is
another scale, the string length, at which the evaporation may be modified
even earlier, due to effects stemming from the finite string size.}  Based
on Page's analysis, roughly 75\% of the original black hole mass might be
emitted in this phase.  As Stephen taught us, a prominent hallmark of this
phase is that the emission is thermal, up to gray-body factors, at any
given time.  Based on this, one may estimate the resulting energy spectrum
of particles in the final state, as well as estimating the ratios of
different kinds of particles produced:  hadrons, leptons, photons, {\it
etc.}  

The decay comes to an end with the \textit{Planck phase}, the final decay
of the Planck-size remnant of the black hole.  This decay is sensitive to
full-blown strongly coupled quantum gravity, and thus its details cannot
yet be predicted.    A reasonable expectation is that the Planck-size remnant
would emit a few quanta with characteristic energy $\sim M_P$.  This end of
the spectrum is where much of the fascination lies:  we can hope for
experimental input on quantum gravity and/or string theory, 
and may see concrete evidence for
the breakdown of spacetime structure, or as Stephen has advocated\cite{Hawking:ra}, even of quantum mechanics.

By putting this all together, we can infer some of the signatures that
would evidence black hole production if it takes place at a future
accelerator.  Decay of a black hole should produce of order $S_{BH}$
primary hard particles, with typical energies given by the Hawking
temperature $T_H$, thus ranging over roughly 100 GeV - 1 TeV.  Creation of
primary particles is essentially democratic among species:  we create an
equal number of each color, spin, and flavor of quark, of each flavor and spin of lepton, of
each helicity state of the gauge bosons, \textit{etc.}  These ratios are then changed
through QCD jet formation, or decays of the primary particles.  For
example, this leads to a rough estimate that we would see five times as
many hadrons as leptons.  Simply the presence of the hard leptons would be
one notable signature.  Moreover, most of the Hawking radiation is isotropic in the
black hole's rest frame, which can't be highly boosted with respect to the
lab frame.  These events would have a high sphericity.  Finally, closer
study may reveal the dipole pattern characteristic of the spindown phase.

So far no one has thought of events based on Standard model physics or any
of its extensions that would mimic these signatures:  if black holes are
produced, their decays should stand out and be discovered.

Nature already provides us with particle collisions exceeding the reach of
the LHC:  cosmic rays hit the upper atmosphere with center-of-mass energies
ranging up to roughly 400 TeV.  We might ask if we could even see black
holes produced by cosmic rays.  Unfortunately, 
the observed flux of ultra-high
energy cosmic rays is believed to consist of either protons or nuclei, and even at
the relevant energies, QCD cross sections dominate the cross section for
black hole production by a factor of roughly a billion.  So most of the
hadronic cosmic rays will scatter via QCD processes before they can make
black holes.  A rough estimate is that these cosmic rays would produce
100 black holes per year over the entire surface of the Earth,
which is clearly too small of a rate to measure\cite{Giddings:2001ih}.

However, it is also believed that there should be a neutrino component of
the ultra-high energy cosmic ray background; this would for example arise
from ultra-high energy protons scattering off the 3K photons in the
microwave background to resonantly produce $\Delta$'s, which produce
neutrinos in their decays\cite{Greisen:1966jv,Stecker:1978ah,Hill:1983mk}.
Neutrinos interact only weakly, and at ultra-high energies 
it turns out that black hole production
is roughly competitive with rates for neutrinos to interact via Standard
Model processes.  Taking existing neutrino flux estimates at face value,
this suggests that neutrinos could produce black holes at rates 
around\cite{Feng:2001ib,DGRT,Anchordoqui:2001ei,Emparan:2001kf}\cite{Giddings:2001ih}\cite{Ringwald:2002vk} 
\begin{equation}
\frac{{\rm several\, black\, holes}}{\rm{yr}\ km^3(H_2O)}\ .
\end{equation}
Detectors that are currently or soon to be operating, such as the HiRes
Fly's Eye, Auger, Icecube, and OWL/Airwatch, are at a level of sensitivity
where they might start to see black hole events, if the assumptions about
the neutrino fluxes are correct.

\section{The future of high energy physics}

High energy physics is a logical extension of a longstanding human quest to
understand nature at an ever more fundamental level.  Once we reach the
Planck scale, things may change; shorter distances than the Planck length
may well not make sense.  However, physics is an experimental subject, and
ultimately we might expect to address the question of shorter-distance
physics experimentally.  However, once black hole production commences,
exploration of shorter distances seems to come to an end.

Specifically, if we want to investigate physics at a distance scale $\Delta
x$ that is shorter than the Planck length, the uncertainty principle tells
us we should scatter particles at
energies $E\sim 1/\Delta x > M_P$.  But if they indeed scatter at distances  $\Delta x$, 
 they will be inside a black hole.  Once a
black hole forms, the outside observer cannot witness the scattering
process directly -- all we see is the Hawking radiation that the black hole
sheds.  Short distance physics is cloaked by black hole formation, and thus
investigation of short distances through high energy physics comes to an
end.

Some of our experimental colleagues might consider this a dismal future.
However, there is another prospect that we can offer them.  As they
increase the energy, they will be making bigger and bigger black holes.  At
some distance scale, these black holes will start to become sensitive to
the shapes and sizes of the extra dimensions, or to other features such as
parallel branes.  When black holes start extending far enough off our brane to
probe these features, their properties, such as their production rate,
their decay spectrum, and other properties, change.  So by doing black hole
physics at increasing energies, experimentalists can reach further off the
brane that we are otherwise confined to, and start to explore the geography
of the extra dimensions of space.  This could certainly continue to yield
exciting experimental discoveries!

\section{Conclusion}

Stephen, I hope you had a happy 60th birthday, but I'm wishing you an even
more exciting 66th birthday!

\bigskip
\bigskip
\centerline{\bf Acknowledgments}
\bigskip

I'd like to thank my collaborators D. Eardley, E. Katz, and S. Thomas for the opportunity to explore these fascinating ideas together.   This work was supported in part by the Department
of Energy under Contract DE-FG-03-91ER40618, and was written up at Stanford University, under partial support from David and Lucile Packard
Foundation Fellowship 2000-13856.

\end{document}